# Collapse of neutral current sheet and reconnection at micro-scales

Shaikhislamov I.F.

Dep. of Laser Plasmas, Institute of Laser Physics, Novosibirsk, 630090, Russia

e-mail: ildar@plasma.nsk.ru

**Abstract:** Reconnection physics at micro-scales is investigated in electron magnetohydrodynamics frame. A new process of collapse of the neutral current sheet is demonstrated by means of analytical and numerical solutions. It shows how a compression of the sheet triggers at scales smaller than ion inertia length an explosive evolution of current perturbation. Collapse results in a formation of intense sub-sheet and then *X*-point structure embedded into the equilibrium sheet. Hall currents associated with this structure support high reconnection rates. Non-linear static solution at scales of electron-skin reveals that electron inertia and small viscosity provide efficient mechanism of field lines breaking. Reconnection rate doesn't depend on the actual value of viscosity, while maximum current is found to be restricted even for space plasmas with extremely rare collisions. Obtained results are verified by a two-fluid large scale numerical simulation.

## 1. Introduction

Magnetic reconnection is a universal process in which magnetic field lines frozen in plasma are forced to break and reform, releasing large amounts of energy in the form of energetic plasma flows. Ideal MHD well describes the physics of reconnection at macro-scales, provided that at some small region a sufficiently high and localized resistivity is introduced [1, 2]. Depending on the size of this region where field lines break, the system might form two basic configurations - Sweet-Parker or Petschek type. Reconnection rates observed in numerous space plasma events support the much faster Petschek model with *X*-point structure of magnetic field. Space plasmas are virtually collisionless and resistivity term in the Ohm's law is orders and orders of magnitude too small to account for observations. Correspondingly, non-ideal region where field lines break should be strongly localized around *X*-point, and involve some new processes as well.

Thus, in the last decade a theory of collisionless reconnection has received considerable attention [3-7]. Its development is based mainly on extensive numerical studies [8-12] using either a Hall-MHD, two-fluid, fully kinetic or hybrid plasma models. Though differing in details most of them give rather similar results. As summarized in [13] these results indicate that the most important non-ideal term in Ohm's law is the Hall term $\boldsymbol{J} \times \boldsymbol{B}/nec$. The latest laboratory experiments also confirm this finding [14]. Electron fluid effectively decouples from ions at scales smaller than ion-inertia length $\lambda_i$ and enforces necessary localization of dissipation region. A number of simulations confirmed that the rate of Hall mediated reconnection doesn't depend on the exact mechanism which finally breaks the frozen-in condition, whether it is electron inertia or grid scale diffusion. This insensitivity is qualitatively explained by the dispersive character of whistler waves, in which the phase velocity increases with decreasing scale size [4, 12, 15]. Consequently, the processes that actually break magnetic field lines have been a focus of research in resent years [17-23]. In a tiny region around the *X*-point of size of the electron-inertia length $\lambda_e$ electron fluid decouples from magnetic field, thus providing, in principal, such collision free mechanism.

Despite the substantial progress made by means of numerical studies, analytical analysis of Hall mediated collisionless reconnection is just developing. A standard approach is to incorporate new effects into the already known MHD solutions [24-27]. Another way is to study reconnection process at scales $< \lambda_i$ in the frame of EMHD, while ignoring ion motion altogether. This approach is used in numerical investigations as well [28, 29]. Detailed treatment of EMHD tearing mode [30] shows that for equilibrium sheets broader than $\lambda_i$ the rates are too small $< 10^{-3} \omega_{ci}$. This suggests a study of other time-spatial modes. In this case, even starting from EMHD, still further approximations have to be carefully chosen. For example, in the detailed analysis [31] a Hall field was rendered to a bipolar form (instead of a quadruple), while in the [32] a bending of reconnecting field lines was excluded.

This paper seeks to understand a physics of magnetic reconnection relevant to the magnetotail. Thus, a standard 2-D configuration based on a Harris sheet is used with no guide field. As a plasma model, a two-fluid description is adopted. No kinetic effects are considered – the approximation which is not easy to justify. Indeed, without the guide field ions are not magnetized at the scales of interest here ($< \lambda_i$).

The ion kinetics could be important and it is needed to reproduce more accurately the rate of ion transport from the *X*-point [33]. However, it doesn't change the rate and the overall picture of reconnection as the comparison of kinetic and Hall MHD codes shows [12, 15, 36]. Concerning the electron kinetics at scales $\sim \lambda_e$, simulations by full particle codes [10, 18] suggest that the non-gyrotropic off-diagonal terms of electron pressure tensor could play a main part in the breaking of field lines. Still, it is fair to say that no widely excepted view exists among the researchers working on this field [21]. Even though fluid dynamics might not be considered valid in a strict sense, as a reference model it could be valuable.

The aim of this work is to demonstrate several analytical solutions relevant to the reconnection physics at scales $< \lambda_i$. A new class of collapse-like solutions is presented. It describes an intense current embedded into the equilibrium sheet which, being driven by the in-plane Hall currents, collapses at the neutral line. Evolution ends by a fast and disruptive restructuring of current sheet from the initially stretched configuration to the typical *X*-point. At scales $< \lambda_e$ a static reconnection solution describing the structure of fields and currents near the *X*-point is derived. It explicitly demonstrates how collisionless reconnection might work without any kinetic effects. A role of collisions is investigated as well. In previous works [12, 15, 16, 34] it was found that a viscosity, rather than resistivity, is the ultimate factor limiting current density. In the present paper a quantitative relation is obtained, which shows that even exceedingly rare collisions are capable to restrict the maximum current density by quite finite albeit large values.

Analysis is carried out in the EMHD frame treating ions as immobile. For non-linear solutions a local approximation along the sheet is employed as well. Because these and other approximations are not strictly justified, obtained results are verified by a direct comparison with the two-fluid numerical simulation. The code used was specifically modified to encompass all scales of interest - from MHD $>> \lambda_i$ to sub-electron-skin $<< \lambda_e$. It should be noted that though very sophisticated codes are in use nowadays, practically all extensive and parametric studies are done with one or other limitation such as the mass ratio or grid size. In the numerical simulation a response of equilibrium system to the boundary perturbation is investigated. It shows triggering of the collapse process and formation of the *X*-point which supports a global quasi-steady reconnection pattern. Characteristic spatial and time scales, current and field structures are found to be in a good agreement with analytical solutions.

The paper is organized as follows. In section 2 models are presented. Dynamic properties of the Hall MHD in application to a plane neutral current sheet are analyzed in section 3. In section 4 this analysis is extended to include finite electron mass and collision effects. The results of numerical simulation are presented in section 5, followed by discussion.

## 2. Models

The problem is restricted to *(x, z)* plane with $\partial/\partial y = 0$. We consider quasi-neutral plasma governed by fluid equations and make use of generalized Ohm's law with the Hall, electron inertia and collision terms included:

$$m\partial \boldsymbol{V_e}/\partial t + m(\boldsymbol{V_e}\nabla)\boldsymbol{V_e} = -e\boldsymbol{E} - e\boldsymbol{V_e} \times \boldsymbol{B}/c - \nabla p_e/n + e\eta \cdot \boldsymbol{J} + v \cdot \nabla^2 \boldsymbol{V_e} \qquad (1)$$

Choosing for characteristic values the asymptotic magnetic field $B_x = B_o$, particle density $n_o$, ion inertia length $\lambda_i = c/\omega_{pi}$, ion cyclotron frequency $\omega_{ci}$ and Alfven velocity equations could be written in the following concise form:

$$\partial n/\partial \tau + \nabla \cdot (n\boldsymbol{V}) = 0, \qquad (2.1)$$

$$n[\partial \boldsymbol{V}/\partial \tau + (\boldsymbol{V}\nabla)\boldsymbol{V}] = \boldsymbol{J} \times \boldsymbol{B} - \nabla P, \qquad (2.2)$$

$$\partial \overline{\boldsymbol{B}}/\partial \tau = \nabla \times [\boldsymbol{V} \times \boldsymbol{B} - \boldsymbol{J} \times \overline{\boldsymbol{B}}/n + \nabla P_e/n] + (1/S_v) \cdot \nabla^2 \overline{\boldsymbol{B}} + (1/S_c - 1/S_v) \cdot \nabla^2 \boldsymbol{B}, (2.3)$$

$$\overline{\boldsymbol{B}} = \boldsymbol{B} - d_e^2 \cdot \nabla^2 \boldsymbol{B}. \qquad (2.4)$$

Here $d_e^2 = m_e/m_i \ll 1$ is the mass ratio. In the chosen units the Lundquist number could be expressed in terms of a collision time as $S_c = \omega_{ce}\tau_{coll}$. Correspondingly, the viscose Reynolds number $S_v$ has been also expressed in terms of an effective collision time $\tau_v$: $S_v = \omega_{ce}\tau_v$, $\tau_v = \lambda_e^2/\nu$. To demonstrate the processes under consideration more clearly, we take the isothermal approximation $T \equiv T_i + T_e = const$. It means that electron pressure effects are excluded from (2.3).

While equations (2) are used for numerical simulation, analysis is based on their reduced EMHD version. In this case (2.1, 2.2) are omitted, plasma velocity is zero, density doesn't vary with time and for simplicity we take it to be constant in space as well. To rewrite (2.3) in a more explicit form, the reversing magnetic field in x-z plane and the out-of-plane main current are expressed through flux function: $\boldsymbol{B}_\perp = -\boldsymbol{e_y} \times \nabla A_y$, $J_y = -\nabla^2 A_y$, while the in-plane currents through Hall field: $\boldsymbol{J}_\perp = -\boldsymbol{e_y} \times \nabla B_y$.

$$d\overline{B}_y/d\tau = -(\boldsymbol{B}_\perp \cdot \nabla)J_y + (1/S_v)\cdot \nabla^2 \overline{B}_y + (1/S_c - 1/S_v)\cdot \nabla^2 B_y, \qquad (3.1)$$

$$d\overline{A}_y/d\tau = (1/S_v)\cdot \nabla^2 \overline{A}_y + (1/S_c - 1/S_v)\cdot \nabla^2 A_y, \qquad (3.2)$$

$$d/d\tau = \partial/\partial \tau - \boldsymbol{J}_\perp \cdot \nabla. \qquad (3.3)$$

There is a simple relation $n \cdot \partial^2(B_y + \nabla \times \boldsymbol{V}_\perp)/\partial \tau^2 = (\boldsymbol{B_o} \cdot \nabla)(\boldsymbol{B_o} \cdot \nabla)B_y$, which follows from (2.3, 2.3) if to ignore collision, electron mass and non-linear terms. Thus, for structures with width $d \ll \lambda_i$ ($z \ll 1$) the Hall current is much larger than the ion velocity as long as the characteristic rates are not much lower than $k_x J_o$. Here $\boldsymbol{B_o} \approx \boldsymbol{e_x} J_o z$ is taken. For equilibrium sheets broader than $\lambda_i$ ($J_o \ll 1$) and correspondingly long wavelengths $k_x \ll 1$ the time-scale on which ion motion could be ignored is much longer than the ion gyro-period.

# 3. Hall dynamics at scales $\lambda_e < d < \lambda_i$

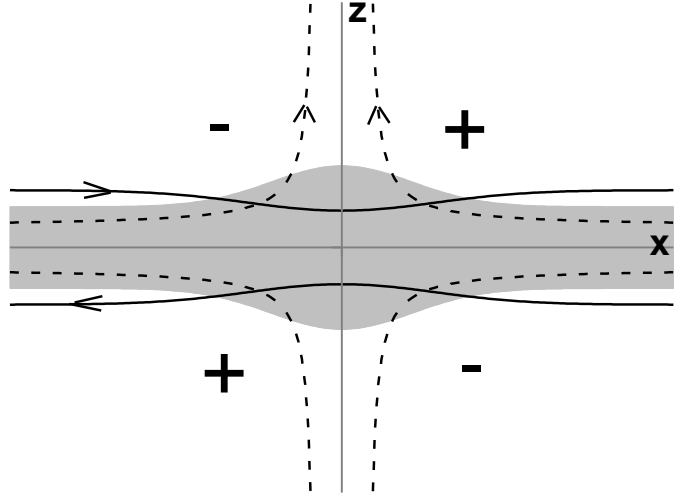

Figure 1. A sketch of the problem. A compressed current sheet is shown as shaded area indicating maximum of $J_y$ at the origin. Plus and minus signs represent a quadruple structure of the Hall magnetic field $B_y$. Reversing magnetic field and Hall currents are indicated by solid and dashed lines respectively.

At the scales of interest here we can ignore in (3) the inertia and collision terms and operate in the frame of collisionless Hall MHD:

$$\partial B_y / \partial \tau = -(\boldsymbol{B}_\perp \cdot \nabla) J_y \tag{4.1}$$

$$\partial A_y / \partial \tau = \boldsymbol{J}_\perp \cdot \nabla A_y = -(\boldsymbol{B}_\perp \cdot \nabla) B_y \tag{4.2}$$

Note that $\boldsymbol{J}_\perp \cdot \nabla B_y \equiv 0$. The geometry of magnetic fields and currents is schematically drawn in fig.1. We will consider thin structures whose width $d$ along $z$ is much smaller than the characteristic length $L$ in $x$-direction. Equations (4) have a peculiar property. Dynamics of the exactly neutral sheets with zero magnetic fields at the neutral line degenerates into a single $x$-dimension. Applying $\partial/\partial z$ to (4.1) and $\partial^2/\partial z^2 + \partial^2/\partial x^2$ to (4.2), and taking equations at $z=0$ where $\boldsymbol{B}=0$, one finds

$$\partial J_x / \partial \tau = J_y \cdot \partial J_y / \partial x \tag{5.1}$$

$$\partial J_y / \partial \tau = -2 J_y \cdot \partial J_x / \partial x + J_x \cdot \partial J_y / \partial x \tag{5.2}$$

Thus, evolution at the neutral line is determined only by a distribution of currents along it. Corresponding *z*-profiles couldn't be found from (5), but the point is they don't influence the evolution at the neutral line. Linearized equations for small perturbations $j_x, j_y$ (here and further denoted by small letters) reduce to

$$\partial^2 j_y / \partial \tau^2 = -2 J_o^2 \cdot \partial^2 j_y / \partial x^2 \qquad (6)$$

Here $J_o = J_y(z=0)$ is the equilibrium current density at the neutral line. Elliptical eq. (6) shows that any perturbation of current at the neutral line grows with increment $\gamma = \sqrt{2} \cdot J_o k$. However, this is not an instability in the usual sense because there is no associated eigenfunction $j_y(z)$. The width of this perturbation in *z*-dimension decreases with the same rate, so in the end it becomes a delta-function at $z = 0$. To see it, let's return to (4) and resolve them along *z*. For periodic perturbation of flux function $a_y = a(z,\tau) \cdot exp(-ikx)$ dynamic equation reads as

$$\partial^2 a / \partial \tau^2 = k^2 B_x^2 (a'' - k^2 a) - k^2 B_x B_x'' a \qquad (7)$$

Here and further the tilde denotes *z*-derivative; $B_x(z)$ is the equilibrium field profile. Evolution of localized perturbation, as governed by (7), could be described in terms of two wave-packets. The one that propagates toward the neutral line experiences steepening of front because of $B_x(z)$ inhomogeneity. To describe the front behavior, it is sufficient to approximate $B_x = J_o \cdot z$ and ignore the small term $k^2 a$ in brackets:

$$\partial^2 a / \partial \tau^2 = J_o^2 k^2 \cdot z^2 a'' \qquad (7.1)$$

The anzats $\theta = \ln z$ transforms (7.1) to a homogeneous form $\partial^2 a / \partial \tau^2 = J_o^2 k^2 \cdot (\partial^2 a / \partial \theta^2 - \partial a / \partial \theta)$, the general solution of which could be described as a Furie series. It shows that while flux perturbation propagates toward negative $\theta$ ($z \to 0$) its amplitude decreases. However, any even derivative at null grows exponentially. Such behavior could be described by a self-similar anzats $a \approx e^{-\gamma_1 \tau} \cdot a(z e^{\gamma_2 \tau})$. Requiring it to be exact solution of (7.1) at least for the first two terms of Taylor expansion $a(z) = a_1 \cdot z^2 + a_2 \cdot z^4$, one gets $\gamma_1 = (2 - \sqrt{2}) \cdot J_o k$; $\gamma_2 = J_o k$.

Non-linearity of Hall-MHD equations transforms this exponential growth into a finite time collapse. For configurations with even $J_y$ and odd $J_x$ eq. (5) could be combined at the null $x = 0$: $\partial^2 \ln(J_y)/\tau^2 = -\partial^2 J_y^2/\partial x^2$. In the local approximation $x \ll L$ with model functions $J_y = J_y(\tau) \cdot \sqrt{1 - x^2/L^2}$, $J_x = -J_x(\tau) \cdot x/L$ solution, obtained for the first time in [35], is

$$J_y(\tau) = \sqrt{2} \cdot J_x(\tau) = J_o / (1 - \tau/\tau_c) \tag{8}$$

Here $\tau_c = L/\sqrt{2} J_o$; $J_o$ is the value of perturbed current at $x = 0$ and $\tau = 0$. Thus, current reaches infinity for a finite time. Note that collapse is triggered only for the reconnection configuration when the *x*-component of electron velocity is directed away at both sides of the null.

Like in the linear case, let's now derive non-linear *z*-profile of currents that corresponds to solution (8). To make the problem tractable we suppose that the width of collapsing structure $d$ is much smaller than its length, and along this large length $L$ we use local approximation. Strictly speaking, such solutions are valid only in the range $z, x \ll L$. However, it could be supposed that overall structure of fields is determined namely in this small region, and that at $z, x \sim L$ it correspondingly adjusts itself. Current experiencing collapse-like behavior could be described as a sheet, whose width goes to zero in a finite time $\tau_c$. Thus, we seek solution in the following self-similar form:

$$A_y = d \cdot A(z/d) \cdot f(x), \quad B_y = B(z/d) \cdot g(x) \tag{9.1}$$

$$d = d_o \cdot (1 - \tau/\tau_c) \tag{9.2}$$

Introducing variable $\theta = z/d$ and substituting (9) into (4), while ignoring all derivatives in respect to $x$ in comparison to $z$, one arrives at

$$\theta \cdot B' = -(\tau_c g'/d_o) \cdot (ff'/gg') \cdot (A'A'' - AA''') \tag{10.1}$$

$$A - \theta \cdot A' = -(\tau_c g'/d_o) \cdot (A'B - (f'g/g'f) \cdot B'A) \tag{10.2}$$

For the Hall field $B_y$ quadruple structure with $g(x) = x/L$, $B(z) = -B(-z)$ should be taken. In the local approximation function $f$ follows from the condition $ff'/gg' = const$; $f(x) = \sqrt{1-(x/L)^2}$. The term $f'g/g'f \sim x^2$ in (10.2) should be ignored and then it follows that $B' = \varepsilon AA''/(A')^2$, $\varepsilon = L \cdot d_o/\tau_c$. After inserting this into (10.1), an integral equation could be written:

$$-J_y \equiv A'' = J_o \exp\left[\int_o^\theta \frac{A'}{A}\left(1 - \varepsilon^2 \cdot \frac{\theta \cdot A}{(A')^3}\right)d\theta\right] \tag{11}$$

Continuity of (11) at $\theta \to 0$ requires that $\varepsilon = \sqrt{2}J_o$, or $\tau_c = L/\sqrt{2}J_o$. At the asymptotic $\theta \to \infty$, where $A \approx J_o \cdot \theta$, current decreases exponentially $A'' \approx J_o \exp(-\theta^2)$. This expression is the first approximation solution. Solution obtained by an iterative numerical integration is shown in fig.2. It exactly corresponds to (8) and gives z-profile of collapsing current.

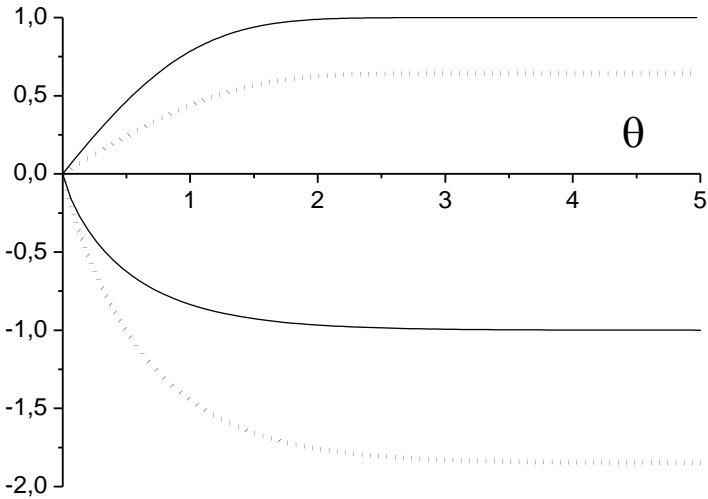

Figure 2. Self-consistent profiles of $b_x$ (solid) and $b_y$ (dotted) components of magnetic field. Positive quadrant – solution (11) with collapse-like variable $\theta = z/(1-\tau/\tau_c)$. Negative quadrant – static solution (13), $\theta = z/d_e$.

Physical interpretation of the collapse is summarized in fig.1. Suppose that, because of a slow compression of the sheet, the main current $J_y$ has a local maximum at the origin and magnetic field lines are bent in as shown. Electrons, moving with a sheared velocity $v_y = -J_y/ne$, stretch these lines out-of-plane, generating the Hall field $B_y$ [4]. Corresponding Hall currents are such that the electron in-plane motion drives the field lines further toward the neutral line and increases the main current $J_y$ at the origin still more. At scales smaller than $\lambda_i$ ions couldn't respond fast enough to changes of fields and the process becomes unstable.

Once the amplitude of perturbation exceeds the equilibrium current ($j_y > J_o$), formed structure becomes so thin that could be considered as embedded into the main current sheet and evolving independently of it. Its outward characteristic is a jump of magnetic field $B_d \equiv \Delta B_x/2 \leq B_o$ which it supports. This value called in literature as the field at the upstream edge of ion dissipation region has been recognized as an important parameter in the Hall mediated reconnection [9]. From fig.2 it is seen that the upstream Hall field $B_h = B_y(\theta \to \infty)$ is smaller but of the same order as $B_d$. The driving Hall current $J_z = B_h/L$ sustains flux flow with the rate $R = J_z B_x \approx V_a B_o \cdot (\lambda_i/L) \cdot (B_d/B_o)^2$. The collapse time is expressed in absolute units as $\tau_c \approx (d_o/V_a) \cdot (L/\lambda_i)$ with $d_o$ being the width of equilibrium sheet. Thus, a non-MHD parameter $\lambda_i/L$ determines an intrinsic rate of Hall processes. Note that Hall MHD is flux conserving ($A_y = const$ at the neutral line) and so far we dealt with the flux transport. Its actual reconnection is considered in the next section.

Above it was demonstrated by two explicit solutions that Hall dynamics has a peculiar tendency to form thin and intense current sub-sheets. However, it contains much more complex behavior. So far we considered either un-localized periodic distribution $exp(ikx)$ or local approximation specified only at small region $kx << 1$. Quite surprisingly, even linearized eq. (6) reveals a number of different and non-trivial scenarios. For arbitrary perturbation of current at the neutral line $j_{yo}(x)$ given at some time exact solution could be written as $j_y = [j_{yo}(x + i \cdot v\tau) + j_{yo}(x - i \cdot v\tau)]/2$, $v = \sqrt{2}J_o$. Because localized perturbation contains many spatial modes, its shape and apparent growth rate changes with time. When localization is relatively "weak" (for example $j_{yo} = sin(kx)/kx$), initial shape, while growing, only slightly changes its form. A bell-like shape $j_{yo} = 1/ch(kx)$ transforms into delta-function $j_y = cos(\tau_k) \cdot ch(kx)/(sh^2(kx) + cos^2(\tau_k)) \to \delta(0)$ in a finite time $\tau_k = kv\tau = \pi/2$. Still more compact perturbation disintegrates into ripples of diminishing scale. For the perturbation $j_{yo} \sim exp(-k^2 x^2)$, which will be used in simulation, solution is $j_y \sim exp(\tau_k^2) \cdot exp(-k^2 x^2) \cdot cos(2kx \cdot \tau_k)$. One can see that at $\tau_k > 1$ the initially smooth shape decays into spikes. Thus, along with contraction, the sub-sheet length decreases too. It suggests a tendency to form point-like, rather than stretched structures, which is a typical feature of Hall mediated reconnection.

### 4. Hall dynamics at scales $< \lambda_e$

It is to be expected that when the width $d$ of collapsing structure decreases down to electron-skin $\lambda_e$ its behavior changes. In this micro-layer the electrons reach high velocities $\sim B_d \cdot V_{ae}$, $V_{ae} = V_a \sqrt{m_i/m_e}$. Here $B_d$ is a normalized value of the upstream field that the intense sub-sheet supports. Because of large current the configuration is no longer neutral $A(0, x, \tau) = -d_e^2 \cdot J_y(0, x, \tau)$, and the reconnection occurs due to finite electron mass. Once again we employ local ($x << L$) and thin sheet ($\partial/\partial x << \partial/\partial z$) approximations. Substituting $A_y = A(z, \tau) \cdot f(x/L)$, $B_y = B(z, \tau) \cdot g(x/L)$ into (3) with the same model functions $g = x/L$, $f = \sqrt{1 - x^2/L^2}$ one obtains

$$\partial \bar{B}/\partial \tau - (B \cdot \bar{B}' - B' \cdot \bar{B} + A'A'' - AA''')/L = 0 \qquad (12.1)$$

$$\partial \overline{A}/\partial \tau - B \cdot \overline{A}'/L = 0 \qquad (12.2)$$

$$\overline{A} = A - d_e^2 \cdot A'', \quad \overline{B} = B - d_e^2 \cdot B'' \qquad (12.3)$$

Equations (12) are valid with accuracy $O(x^2/L^2)$. Collision terms are omitted for the time being. Evolution at scales $< \lambda_e$ could be described in the following way. Because of electron inertia, flux function $A$ and Hall field $B$ behave smoothly, while the effective values $\overline{A}, \overline{B}$ develop progressively larger gradients near null. Any changes are restricted to a small area, while everywhere else magnetic fields reach quasi-stationary profiles. These profiles could be derived from static eqs. (12) with a given flux inflow rate (or parallel electric field) $\partial A/\partial \tau = R$:

$$d_e^2 BB'' - B_x^2 + B_d^2 - A \cdot J_y - (d_e \cdot B')^2 = 0 \qquad (13.1)$$

$$R + (B_x - d_e^2 B_x'') \cdot B/L = 0 \qquad (13.2)$$

where $B_x = -A'$ and $J_y = -A''$. Eq. (13.1) is derived by integration of (12.1) with the constant $B_d = B_x(z \to \infty)$. Near the null where $A \approx -d_e^2 \cdot J_y$ the largest in (13.1) are the two last terms which are traced in the electron momentum equation (1) to the production $eV_e \times B/c$ and convection $m(V_e \nabla)V_e$ terms respectively. The Hall field and current follow as $B \approx -B_x$, $J_x(z) \approx J_y(z)$. Using this, (13.2) could be expressed near the null as a simple first order differential equation for $B_x$:

$$d_e \cdot B_x'/B_d = -\sqrt{(B_x/B_d)^2 - 1 - (RL/B_d^2) \cdot ln(B_x/B_d)^2} \qquad (14)$$

Approximate solution is $B_x \approx -\sqrt{RL} \cdot \theta \cdot \sqrt{-2ln\theta}$, $\theta = z/d_e$. It has weak singularity at $z = 0$ where its derivative logarithmically diverges: $J_y \approx -\sqrt{RL}/d_e \cdot \sqrt{-2ln\theta}$. Exact numerical integration of (13) is shown in fig.2. It coincides with (14) at $\theta << 1$, while the asymptotic behavior at $\theta >> 1$ requires that a relation between the reconnection rate and the upstream field is $R \approx 2 B_d^2 / L$. For upstream Hall field the value $B_h \approx 2 \cdot B_d$ follows.

Time behavior of currents at the null could be easily found from dynamic equations (12). When flux increases with the same rate everywhere $\partial A/\partial \tau = R$, conservation of canonical flux $A_y - mcV_{ey}/e = const$ dictates that current at the neutral line increases as $J_y \sim neV_{ae} \cdot \tau/\tau_c$ with characteristic time $\tau_c = d_e \cdot \sqrt{L/R}$. Note that this is just the time needed for electrons to pass the skin length: $\tau_c \approx d_e/J_z$. Self-consistent convection that acts via the term $m(V_e \nabla)V_e$ confines current increase to a progressively smaller region, so as the main dynamic feature is extremely fast exponential growth of the current gradient: $J_y'' \sim exp(\tau^2/\tau_c^2)$. This kind of development is fully consistent with the static solution (14), which obviously is the asymptotic state of collisionless evolution.

To resolve any singularities and find out the final state of intense collapsing structure, it is necessary to take into account collisions. It is sufficient to investigate the static reconnection solution of (3). When electron mass is ignored the resistive sheet width follows as $d_c = S_c^{-1}\sqrt{L/R} \approx (S_c J_z)^{-1}$ [24]. In the case of rare collisions $d_c << d_e$ we need to account for them only in a small region $<< d_e$ near the X-point while outside it collisionless static solution (14) could be used. Eq. (12.2) with the collision terms retained could be rewritten as

$$\frac{\partial J_y}{\partial \theta} = S_v R \cdot e^{-K_v \cdot \chi} \cdot \int_o^\theta \left[ 1 - \frac{B \cdot B_x}{RL} + \frac{J_y}{S_c R} \right] \cdot e^{K_v \cdot \chi} \cdot d\theta \qquad (15)$$

where $K_v = d_e/d_v >> 1$ is a ratio of electron-skin to viscose scale $d_v = S_v^{-1}\sqrt{L/R}$; $\chi(\theta) = \int_o^\theta \frac{B_y}{B_d} \cdot d\theta$; $\theta = z/d_e$. At the right side of (15) it is sufficient to use dependences given by (14). Characteristic gradient scale follows, from a condition $K_v \cdot \chi(\theta) = 1$, as a geometrical mean value of viscose and electron-skin scales:

$d = \sqrt{d_v d_e} \approx \sqrt{d_e/(S_v J_z)}$; $d_v \ll d \ll d_e$. Now, current derivative at the null behaves continuously: $J_y' = -(S_v R/d_e^2) \cdot z$, while at $z > d$ it is given by collisionless solution $J_y \approx -\sqrt{RL}/d_e \cdot \sqrt{-2\ln\theta}$. The maximum value could be found in the first approximation by taking this expression at $z = d$:

$$J_y \approx \sqrt{RL}/d_e \sqrt{\ln\left(S_v d_e \sqrt{\frac{R}{L}}\right)} \tag{16}$$

Thus, the maximum current density is an extremely weak function of viscosity. Only at extremely low viscosity in comparison to the resistivity $S_v/S_c < (d_e/d_c) \cdot exp(-d_e^2/d_c^2)$ current at the neutral line reaches its usual (and very large) resistive MHD value $J_y = S_c R$. Physical reason of the viscosity importance is clear – at scales $\ll \lambda_e$ current gradients are very large.

Calculating the energy balance, one can see that the magnetic energy that flows to the neutral line $P_{in} = L \cdot (J_z \cdot B_x^2)_{z=\infty} \sim B_d^3$ is converted into the kinetic energy of the out-flowing electrons $P_{out} = d_e^2 \int dz \cdot J_x (J_x^2 + J_y^2)/2 \sim B_d^3$. Another interesting feature is that the distribution of electrons on kinetic energies inside the electron-skin is described by the exponential function $dE_k/dN \sim exp[-E_k/(E_{ae} \cdot B_d^2)]$, $E_{ae} = B_o^2/(4\pi n)$ with faster electrons near the neutral line. Cut off depends on the viscosity and could be substantially larger than $E_{ae}$.

Obtained solution describes a steady state reconnection with electron inertia and a small electron viscosity. Due to inertia the bulk of electrons in the layer $\leq \lambda_e$ moves with electron-Alfven speed. In the limit of vanishing viscosity electron velocity and current near the null could be arbitrary large, but it doesn't impose any problem because a number of such electrons is exponentially small. In other words, finite electron mass transforms current profile in such a way that not only the net current $\int J_y dz \sim B_d$ is finite, but the equivalent of energy $m_e \int J_y^2 dz \sim B_d^2 \cdot E_{ae}$ as well.

## 4. Numerical simulation

In this section the above analytical analysis is verified against the results of two-fluid 2D numerical simulation of equations (2). Code implementation is the same as in [35]. Because of necessity to cover a five order span of spatial scales, an additional non-uniform mesh at $z \leq \lambda_i$ was employed. For the equilibrium state a Harris current sheet was taken with magnetic pressure being balanced with a thermal pressure. Its width was chosen to be in the MHD range $d_o = 5 \cdot \lambda_i \gg \lambda_i$ so as to ensure a clear division between ion and electron dominated regions. The realistic mass ratio was taken $d_e^2 = 1/1836$. The minimum mesh size $\approx 10^{-4} \lambda_i$ dictated correspondingly small time step $\approx 5 \cdot 10^{-4}/\omega_{ci}$ which was reduced still more at later times as the intensity of currents increased. Evolution of the sheet is initiated by imposing a small-amplitude boundary perturbation, which is a forced reconnection problem. Instead of periodic, a localized perturbation was chosen, described by the asymptotic rate $R_f$ of forced flux inflow and the rise time $\tau_f$:

$$\partial A_y(z_{box}, x, \tau)/\partial \tau = R_f \cdot exp(-x^2/L^2) \cdot \xi(\tau/\tau_f) \tag{17}$$

Rather than switch off the flux inflow at some time and see what maximum reconnection rate the system achieves, we expect it to reach the rate equal to $R_f$. For this the function $\xi(\theta) = \theta^2/(1+\theta^2)$ was used with $L = 20\lambda_i$ and $\tau_f = 50/\omega_{ci}$. To avoid unnecessary boundary effects, the system size was taken to be $z_{box} = 15\lambda_i$, $x_{box} = 70\lambda_i$. Free boundary conditions were imposed at the *x*-ends so plasma could flow through them. The Lundquist number was chosen to be sufficiently large $S_c > 10^6$ so as to exclude spontaneous development of the tearing mode. It was checked that without boundary perturbation the system is stable.

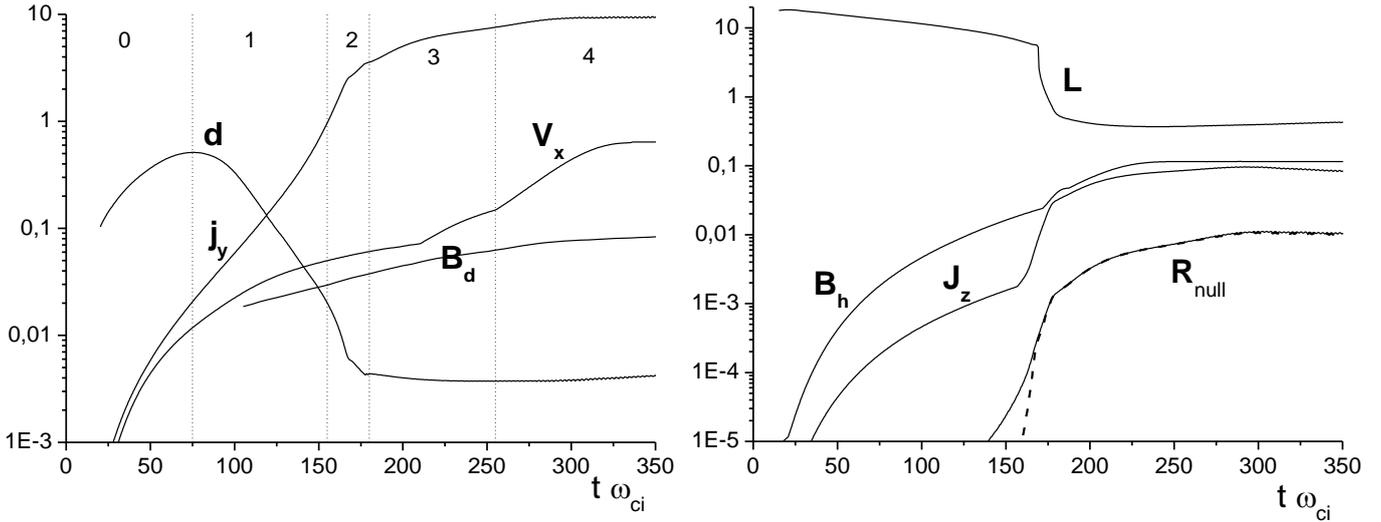

**Fig. 3-a** Behavior of current perturbation at the neutral line $j_y$, its half-width $d$, upstream field $B_d$ which is supported by this current and maximum out-flow velocity of ions $V_x$. The following phases of development are marked: 0 – build up of perturbation, 1 – collapse, 2 – sub-electron skin evolution, 3 – formation of X-point and accumulation of reconnected flux, 4 – tearing instability. Parameters of simulation: $J_o = 0.2$, $R_f = 0.01$, $L = 20$, $S_v = 10^6$, $S_c = \infty$.

**Fig. 3-b** Behavior of values related to Hall dynamic – maximum out-of-plane field $B_h$, maximum in-flow current $J_z$, half-length of the sub-sheet $L$ and reconnection rate $R_{null}$. Dashed line – viscose term $(\lambda_e^2 J_y'')/S_v$ at the null.

In fig.3 a typical example of time-behavior of various values is shown for the case $R_f = 0.01$. Here $j_y$ is the current perturbation at the origin; $d$ is the half-width of $j_y$ profile across the neutral line and $L$ - along it; $B_h$ was measured as a maximum of $B_y$ component in the calculation box, while method of deriving the value of $B_d$ is demonstrated in fig.4 below. Evolution could be divided into several phases. Initial phase marked by zero is a simple MHD development caused by boundary perturbation. At this stage the ion flow builds up (presented by the maximum out-flow velocity $V_x$). At phases 1-3 it slowly reaches asymptotic pattern with the boundary inflow velocity $V_z = -R_f$. This flow slightly compresses and bends the current sheet. Corresponding perturbation of the main magnetic field $B_d$ is shown on the same figure and closely follows ion dynamic. On the other hand, behavior of the current perturbation $j_y$ and its half-width $d$ at phases *1, 2* is altogether different from ion dynamic. It clearly shows formation and collapse of intense sub-sheet. In time interval $75 < \tau < 165$ a best fit for the current rise is $j_y \sim exp\left((\tau+1.2)^2/70.5^2\right)$, like predicted for perturbations localized as $exp(-x^2/L^2)$. Moreover, the characteristic time appears to be in a very good agreement with the one derived in section 3: $\tau_c = L/\sqrt{2}J_o \approx 71$. Additionally to check analysis, it was verified that small periodic perturbation $cos(x/L)$ grows as predicted $j_y \sim exp(\tau/\tau_c)$ with the same $\tau_c$.

Evolution of the sub-sheet at sub-electron inertia scales $< d_e \approx 0.023$, until it reaches collision width determined by viscosity, proceeds much faster. In fig.3-b behavior of several other parameters is shown. From them all a most dramatic change exhibits the second derivative of current at the null. At phase 2 it could be very well approximated by $J_y'' \sim exp\left((\tau-141)^2/11.3^2\right)$, which agrees with the result of section 3, along with the estimate of characteristic time $\tau_c = d_e/J_z \approx 14.3$ (as $J_z$ the maximum value measured at time $\tau = 150$ is taken, see fig. 3b). Z-profile of the sub-sheet after the collapse is demonstrated in fig.4. For comparison, by dotted line the profile derived from (13) is drawn, exactly the same as in fig.2. There is quite a good agreement at $z < d_e$. Straight dashed line illustrates how the upstream value of perturbed magnetic field $B_d$ is determined. In the inserted panel profile of current $J_y$ at scales far below $<< d_e$ is presented. It confirms a slow logarithmic increase down to collision dominated width $\approx 5 \cdot 10^{-4}$. This width is close to analytical estimate $d = \sqrt{d_e/(S_v J_z)} \approx 7.4 \cdot 10^{-4}$ (with $J_z$ measured at $\tau = 190$). The viscose number for this run was $S_v = 10^6$, while Lundquist number $S_c$ didn't produce any effect and was chosen to be infinite.

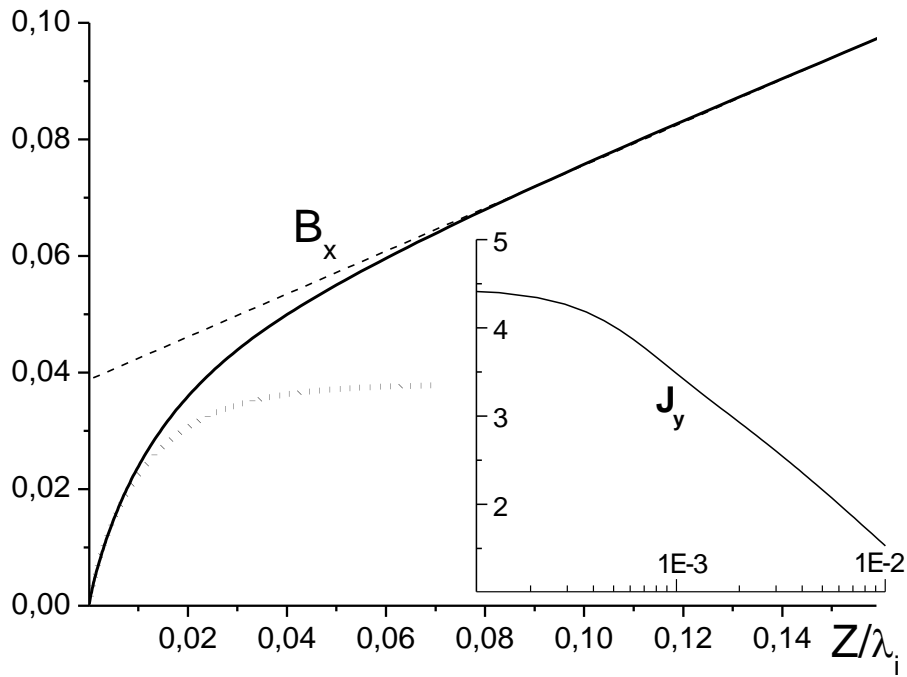

**Fig. 4** Profile of the main field $B_x$ at electron-skin scale. Solid line – simulation; dotted line – solution of (13); dashed line – linear fit of the field outside the sub-sheet. Its crossing with abscissa gives upstream value $B_d$. **Inserted panel:** Current profile at sub-electron-skin scale.

Evolution of the sub-sheet at phase 2, until it reaches quasi-static state, involves drastic change of its 2D structure as well. During short time from $\tau \approx 170$ to $\tau \approx 180$ the length $L$ of sub-sheet suddenly drops by order of magnitude (fig.3-b). What really happens is clearly seen from fig.5, where 2D plots of the main current and Hall field are drawn at times before and after this event. Namely, the stretched current structure splits into the *X*-point with four separate wings (only one quadrant is shown at the picture). At later times the wings gradually extend further away from the null forming a typical cross shape. This extremely short (only several ion cyclotrons periods!) restructuring of current sheet could be well associated with the so called disruption events. In the similar EMHD simulation [28] qualitatively the same picture was reported, with the exception that current was seen to split into two *X*-points located at system boundaries (probably because of periodic conditions used) and connected with extended wings. Also, like in [28], intensive whistler oscillations and multiple small islands were observed as an after-effect of disruption. This is demonstrated in fig.6 where current profiles along the neutral line are shown at the two moments of time. The way how sub-sheet erodes and forms a sharp spike well supports the arguments given in section 3.

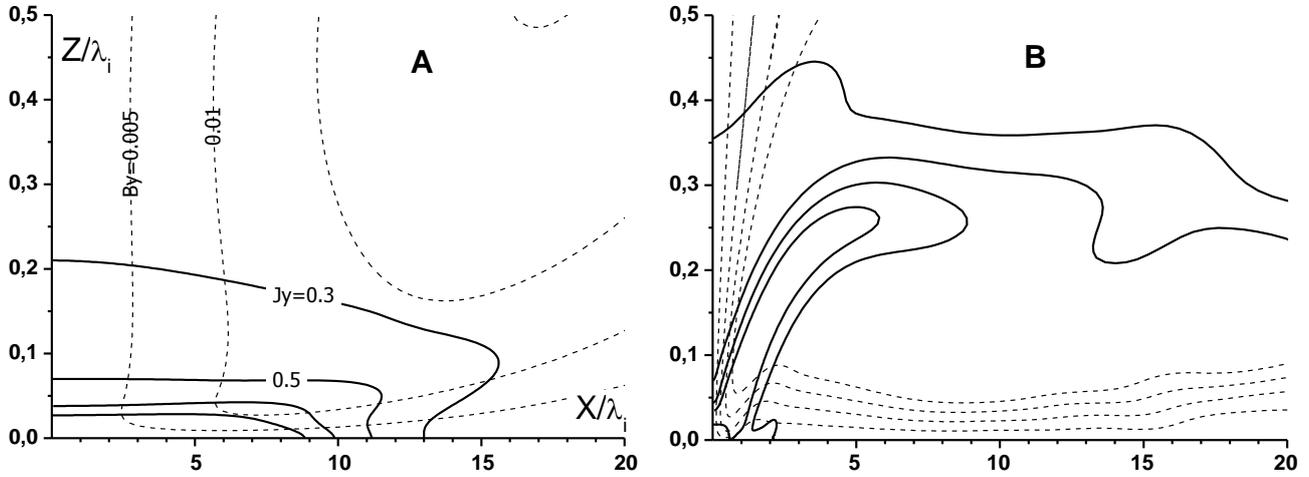

**Fig. 5** Contour plots of current $J_y$ and Hall field $B_y$ at time $\tau = 160$ (A) and $\tau = 190$ (B). Isoclines of current are incremented by 0.2 starting from 0.3. Isoclines of Hall field are incremented by 0.005 starting from 0.005.

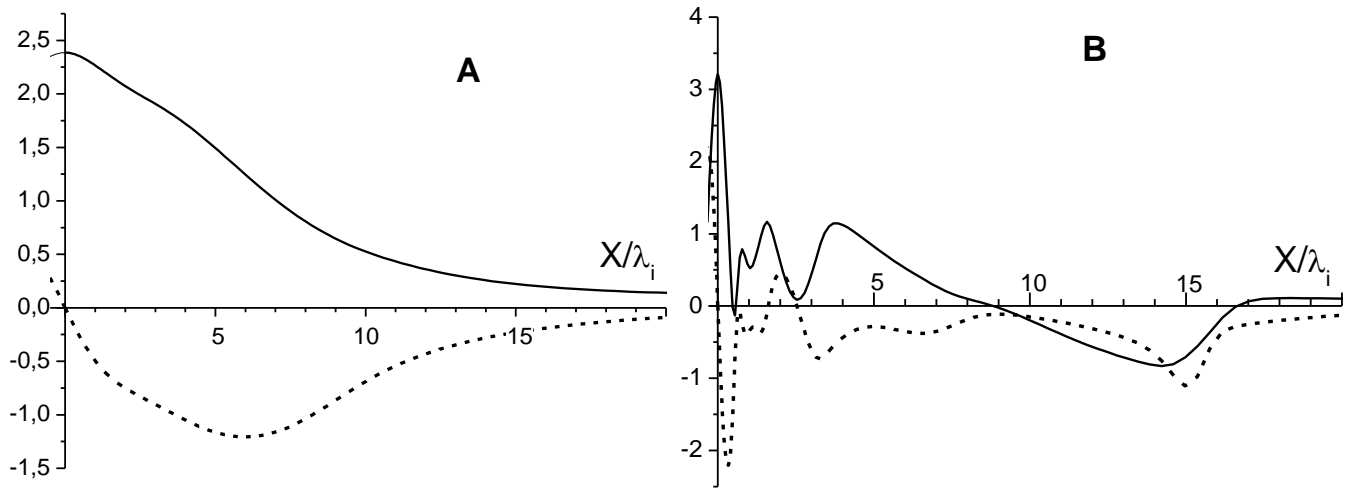

**Fig. 6** Distribution of main current $J_y$ (solid curve) and out-flow Hall current $J_x$ (dash curve) along the neutral line at time $\tau = 165$ (A) and $\tau = 175$ (B).

At the end of collapse the reconnection rate $R_{null} = \partial A(z, x = 0)/\partial t$ approximately reaches the flux inflow rate $R_f$ imposed at the boundary. In fig.7 behavior $R_{null}(\tau)$ is shown for several different values of $R_f$.

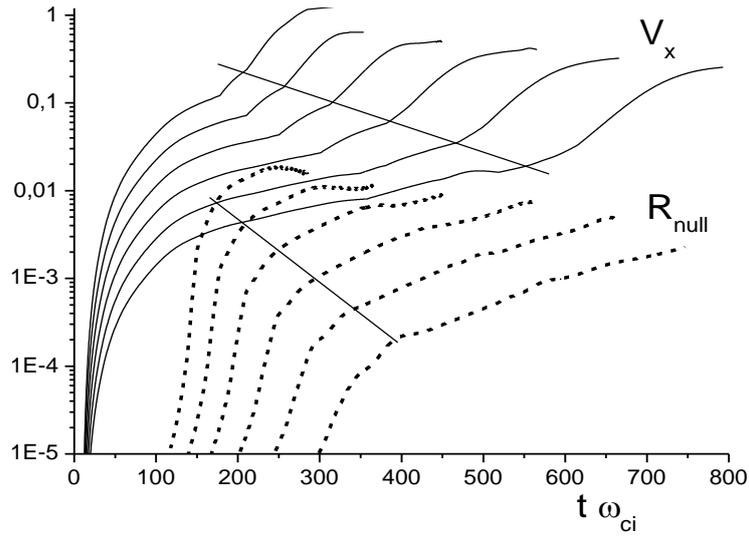

**Fig. 7** Reconnection rate $R_{null}$ (dash lines) and maximum out-flow velocity $V_x$ (solid lines) versus time for several values of flux inflow at the boundary. First curve - $R_f = 0.02$, for each next one $R_f$ is twice smaller. Straight lines indicate moments of collapse end and moments of tearing instability onset.

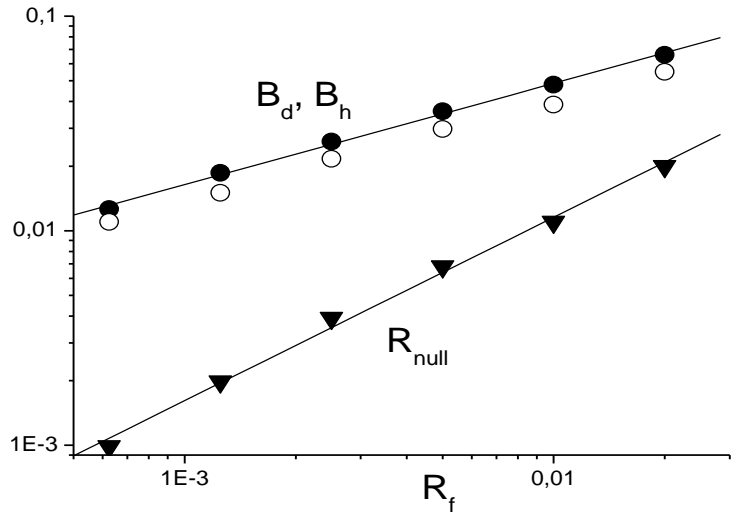

**Fig. 8** Dependences on the amplitude of boundary perturbation: reconnection rate $R_{null}$ (black triangles), main field $B_d$ (white circles), Hall field $B_h$ (black circles).

If measured at characteristic moments (as marked on fig.7) $R_{null}$ scales linearly with $R_f$, while the upstream values of Hall and main field $B_h, B_d$ scale as a square root, as shown in fig.8. This confirms that Hall mediated reconnection is essentially non-linear effect. It is interesting to note that the two-stage behavior of the reconnected flux with a weak increase following the initial surge reported in the latest multi-code numerical investigation of forced reconnection [36] is also well seen in fig.7. The system behavior was found to be totally insensitive to the collision terms (as long as they are not very large). The only value that depends on it is the current density at the null. It is shown in fig.9, as well as calculation by (16) with corresponding parameters. One can see that results differ only by a minor constant shift and have the same very weak scaling.

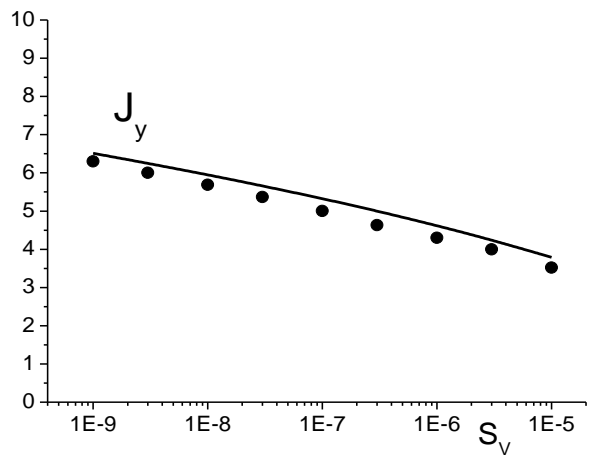

**Fig. 9** Dependence of maximum current density on the viscose number. Solid curve shows calculation by formula (17).

Investigation of a large-scale and long-time behavior of the system isn't the purpose of this paper. However, some remarks are worth mentioning. Once the Hall currents form *X*-point micro-structure which supports fast reconnection of field lines in a small region near the null, it is to be expected that the tearing instability will dominate large-scale ion dynamics. At the phase "3" of development (see fig.3-a) accumulation of reconnected flux courses acceleration of ions. At this time the location of maximum inflow velocity $V_z$ shifts from the boundary $z = Z_{box}$ to the region inside of the main current sheet $z \sim \lambda_i$ where it significantly exceeds $R_f$. From fig.7 one can clearly see that at the last phase "4" ion out-flow velocity increases exponentially with the same increment regardless of the amplitude $R_f$ of initiating perturbation.

## 5. Discussion and Conclusions

Numerous numerical studies report, without much discussion, that reconnection phase always appears as a sharp jump after a long developmental evolution. Solutions obtained in this work demonstrate how a slow bending of opposite magnetic field lines triggers at scales $< \lambda_i$ the collapse of current perturbation and subsequent formation of intense sub-sheet with associated Hall currents which are responsible for a sudden increase of reconnection rate. Characteristic time of this process is determined mainly by the length of region where field lines are bent-in and by the width of equilibrium sheet (or equilibrium current density). Evolution of the sub-sheet has tree distinct phases – collapse, formation of sub-electron-skin layer and X-point restructuring.

The same tendency to form rapidly shrinking sub-sheets was found and described in [27, 34]. Mathematical model based on the incompressible MHD equations with electron inertia included is the same as (3) if $B_y$ is replaced by a streamfunction $\varphi$ and $\bar{B}_y$ by a vorticity $\nabla^2 \varphi$. However, without the Hall term this model is incomplete as the scales involved are much smaller than $\lambda_i$.

One of the going-on discussions in the literature concerns a role of the upstream value of magnetic field $B_d$ at the edge of ion dissipation region [9, 16]. We find it to be relevant for the understanding of reconnection process at micro-scales. Results of this work suggest that this field is directly related to the intense sub-sheet. Obtained solutions yield, as it was derived qualitatively in previous works [4, 24], that Hall mediated reconnection rate is $R \sim (\lambda_i / L) \cdot B_d^2$. Extensive numerical study [9] showed that $B_d$ depends on initial conditions and prior evolution, while the length $L$ of boundary between electron and ion dominated regions is determined by the in-flow and out-flow velocities of ions. In the present study of reconnection in a small region around the X-point where electron dynamics dominates, namely those characteristic values appear as open parameters in EMHD solutions.

If the collision dissipation region is smaller than $\lambda_e$, which is the case for space plasmas, it was doubted that other fluid terms, which are electron inertia terms $m\partial \boldsymbol{V_e}/\partial t$, $m(\boldsymbol{V_e}\nabla)\boldsymbol{V_e}$, could support reconnection. Indeed, in quasi-static case the first one should be neglected while the second is zero at the null. Also, arguments based on whistler wave dispersion predict that reconnection rate in this case should depend on the electron mass, since at scales $<\lambda_e$ whistler waves are no longer quadratic. Hence, kinetic effects, including off-diagonal pressure tensor terms, have been a main focus of research. However, even if the inflow electron velocity $V_{ez}$ is zero at null and, for the given electric field $E_y$, current carrying component $V_{ey}$ is formally infinite, solution could be still correct as long as energy and momentum conservation isn't violated (which is automatic if full momentum equation is used). Static reconnection solution derived in section 4 is, in a way, rather obvious. In the skin layer electrons transport mostly current instead of flux $E_y = -(\boldsymbol{V_e}\nabla)(A_y + \lambda_e^2 J_y) \approx -\lambda_e^2(\boldsymbol{V_e}\nabla)J_y$. It means that converging flow $V_{ez} \sim z$ piles up current $J_y \sim \ln(z)$ while magnetic field remains monotonic.

It was pointed out before [12, 15] that only viscosity could balance whistler mediated reconnection. In this work a smallest possible scale in the problem determined by viscosity was derived. It was found that, due to a very weak dependence $J_{y,max} \sim \sqrt{\ln(\tau_{coll})}$, collisions effectively restrict maximum current density. While dissipation doesn't affect the reconnection rate, it still plays a fundamental role by making the change of topology truly irreversible. Indeed, without dissipation energy conversion proceeds in a much more complex way through secondary instabilities and turbulent cascade [37].

The essence of presented results could be summarized as follows. By means of collapse process the electron fluid forms such a flow that transports flux with the same rate through ion ($<\lambda_i$) and electron ($<\lambda_e$) skins right to the *X*-point. Around some tiny region around the *X*-point current piling up gives rise to the viscose dissipation and irreversible flux reconnection.

There are a number of questions that needs a further study. Though simple arguments given in section 3 show that Hall dynamics has an intrinsic tendency to form *X*-point structures, a two-dimension solution that describes the cross-shaped current configuration is needed. Also, it is interesting to see how the normal component of magnetic field typically present in equilibrium current sheets affects the collapse process.

**Acknowledgements.** Helpful discussions with Zakharov Yu.P. are appreciated.